\documentclass[10pt,twocolumn]{article}%
\usepackage{setspace}
\usepackage{color}
\usepackage{enumerate} 
\usepackage{epsfig} 
\usepackage{amssymb} 
\usepackage{fancyhdr}
\usepackage{amsmath}
\usepackage{graphicx}
\usepackage{natbib}
\usepackage{multirow}
\usepackage{rotating}
\usepackage{verbatim}
\usepackage[left=0.5in,top=0.75in,right=0.5in,bottom=0.5in]{geometry}

\newcommand{\negspace}{\!\!\!\!\!\!\!\!\!\!\!\!}
\usepackage{url}

\begin{document}

\title{ {\tt selscan}: an efficient multi-threaded program to perform EHH-based scans for positive selection}
\author{Zachary A. Szpiech$^{1,*}$ and Ryan D. Hernandez$^{1,2,3}$}

\maketitle
\thispagestyle{fancy}
\begin{flushleft}
	{\it $^1$ Department of Bioengineering and Therapeutic Sciences, University of California, San Francisco, San Francisco, CA, USA}\\
	{\it $^2$ Institute for Human Genetics, University of California, San Francisco, San Francisco, CA, USA}\\
	{\it $^3$ Institute for Quantitative Biosciences (QB3), University of California, San Francisco, San Francisco, CA, USA}\\
	$^*$ zachary.szpiech@ucsf.edu
\end{flushleft}

\abstract{Haplotype-based scans to detect natural selection are useful to 
identify recent or ongoing positive selection in genomes.  As both real and simulated 
genomic datasets grow larger, spanning thousands of samples and millions of markers, there is a 
need for a fast and efficient implementation of these scans for general use. 
Here we present {\tt selscan}, an efficient multi-threaded application 
that implements Extended Haplotype Homozygosity (EHH), Integrated Haplotype 
Score (iHS), and Cross-population Extended Haplotype Homozygosity (XPEHH).  
{\tt selscan} accepts phased genotypes in multiple formats, including TPED, and performs extremely well on both simulated and real data and over an order of 
magnitude faster than existing available implementations. It 
calculates iHS on chromosome 22 ($22,147$ loci) across $204$ CEU haplotypes in $353$s on 
one thread ($33$s on $16$ threads) and calculates XPEHH for the same data relative to $210$ YRI
haplotypes in $578$s on one thread ($52$s on $16$ threads).  Source code and binaries 
(Windows, OSX and Linux) are available at {\tt https://github.com/szpiech/selscan}.}

\section{Introduction}

Extended Haplotype Homozygosity (EHH) \cite[]{SabetiEtAl02}, Integrated Haplotype 
Score (iHS) \cite[]{VoightEtAl06}, and Cross-population Extended Haplotype 
Homozygosity (XPEHH) \cite[]{SabetiEtAl07} are statistics designed to use phased genotypes to identify putative regions
of recent or ongoing positive selection in genomes.  They are all based on the model of 
a hard selective sweep, where a {\it de novo} adaptive mutation arises on a haplotype
that quickly sweeps toward fixation, reducing diversity around the locus.  If selection is strong enough, this occurs
faster than recombination or mutation can act to break up the haplotype, and thus a signal of high 
haplotype homozygosity can be observed extending from an adaptive locus. 

As genetics data sets grow larger both in number of individuals and number of loci,
there is a need for a fast and efficient publicly available implementation of these statistics. Below we 
introduce these statistics and provide concise definitions for their calculations.  We 
then evaluate the performance of our implementation, {\tt selscan}.

\subsection{Extended Haplotype Homozygosity}

In a sample of $n$ chromosomes, let $\mathcal{C}$ denote the set of all possible
distinct haplotypes at a locus of interest (named $x_0$), and let 
$\mathcal{C}(x_i)$ denote the set of all possible distinct haplotypes 
extending from the locus $x_0$ to the $i^{th}$ marker either upstream or 
downstream from $x_0$.  For example, if the locus of interest $x_0$ is a 
biallelic SNP where $0$ represents the ancestral allele and $1$ represents the 
derived allele, then $\mathcal{C} := \{0,1\}$.  If $x_1$ is an immediately 
adjacent marker, then the set of all possible haplotypes is 
$\mathcal{C}(x_1) := \{11,10,00,01\}$.

EHH of the entire sample, extending from the locus $x_0$ 
out to marker $x_i$, is calculated as 
\begin{equation}\label{eq:ehh}
EHH(x_i) = \sum_{h \in \mathcal{C}(x_i)} \frac{{n_h \choose 2}}{{n \choose 2}},
\end{equation}
where $n_h$ is the number of observed haplotypes of type $h \in \mathcal{C}(x_i)$.

In some cases, we may want to calculate the haplotype homozygosity of a 
sub-sample of chromosomes all carrying a `core' haplotype at locus $x_0$. Let 
$\mathcal{H}_c(x_i)$ be a partition of $\mathcal{C}(x_i)$ containing all 
distinct haplotypes carrying the core haplotype, $c \in \mathcal{C}$, at $x_0$ and 
extending to marker $x_i$.  Note that
\begin{equation} 
\mathcal{C}(x_i) = \bigcup_{c \in \mathcal{C}}\mathcal{H}_c(x_i).
\end{equation}

Following the example above, if the derived allele (1) is chosen as the core haplotype,
then $\mathcal{H}_1(x_1) := \{11,10\}$.  Similarly, if the ancestral allele is the core
haplotype, then $\mathcal{H}_0(x_1) := \{00,01\}$

We calculate the EHH of the chromosomes carrying the core haplotype $c$ to marker $x_i$ as
\begin{equation}\label{eq:ehh-core}
EHH_c(x_i) = \sum_{h \in \mathcal{H}_c(x_i)} \frac{{n_h \choose 2}}{{n_c \choose 2}},
\end{equation}
where $n_h$ is the number of observed haplotypes of type $h \in \mathcal{H}_c(x_i)$ and 
$n_c$ is the number of observed haplotypes carrying the core haplotype ($c \in \mathcal{C}$). 

\subsection{Integrated Haplotype Score}

iHS is calculated by using Equation \ref{eq:ehh-core} to track the decay of haplotype homozygosity
for both the ancestral and derived haplotypes extending from a query site.
To calculate iHS at a site, we first calculate the integrated haplotype 
homozygosity (iHH) for the ancestral ($0$) and derived ($1$) haplotypes 
($\mathcal{C} := \{0,1\}$) via trapezoidal quadrature.
\begin{align}\label{eq:ihh}
iHH_c =& \notag\\
&\negspace\sum_{i = 1}^{|\mathcal{D}|} \frac{1}{2}\left(EHH_c(x_{i-1}) + EHH_c(x_i)\right)g(x_{i-1},x_i) + \notag\\
&\negspace\sum_{i = 1}^{|\mathcal{U}|} \frac{1}{2}\left(EHH_c(x_{i-1}) + EHH_c(x_i)\right)g(x_{i-1},x_i),
\end{align}
where $\mathcal{D}$ is the set of markers downstream from the current locus 
such that $x_i \in \mathcal{D}$ denotes the $i^{th}$ closest downstream 
marker from the locus of interest ($x_0$). $\mathcal{U}$ and $x_i \in \mathcal{U}$ are defined similarly 
for upstream markers. $g(x_{i-1},x_i)$ gives the genetic distance between two 
markers.  The (unstandardized) iHS is then calculated as
\begin{equation}
\ln\left(\frac{iHH_1}{iHH_0}\right).
\end{equation}
Note that this definition differs slightly from that in \cite{VoightEtAl06}, where 
unstandardized iHS is defined with $iHH_1$ and $iHH_0$ swapped.

Finally, the unstandardized scores are normalized in frequency bins across the 
entire genome.
\begin{equation}
iHS = \frac{\ln\left(\frac{iHH_1}{iHH_0}\right) - E_p\Big[\ln\left(\frac{iHH_1}{iHH_0}\right)\Big]}{SD_p\Big[\ln\left(\frac{iHH_1}{iHH_0}\right)\Big]},
\end{equation}
where $E_p\Big[\ln\left(\frac{iHH_1}{iHH_0}\right)\Big]$ and 
$SD_p\Big[\ln\left(\frac{iHH_1}{iHH_0}\right)\Big]$ are the expectation and 
standard deviation in frequency bin $p$.

In practice, the summations in Equation \ref{eq:ihh} are truncated once 
$EHH_c(x_i) < 0.05$.  Additionally with low density SNP data, if the physical 
distance $b$ (in kbp) between two markers is $> 20$, then $g(x_{i-1},x_i)$ is 
scaled by a factor of $20/b$ in order to reduce possible spurious signals 
induced by lengthy gaps.  During computation if the start/end of a chromosome 
arm is reached before $EHH_c(x_i) < 0.05$ or if a gap of $b > 200$ is 
encountered, the iHS calculation is aborted for that locus.  iHS is not 
reported at core sites with minor allele frequency $< 0.05$.  In {\tt selscan}, the 
EHH truncation value, gap scaling factor, and core site MAF cutoff value are
all flexible parameters definable on the command line.

\subsection{Cross-population Extended Haplotype Homozygosity}

To calculate XPEHH between populations $A$ and $B$ at a marker $x_0$, we first calculate iHH for each population separately, integrating the EHH of the entire sample in the population (Equation \ref{eq:ehh}).
\begin{align}\label{eq:xpihh}
iHH =&\notag\\
&\negspace\sum_{i = 1}^{|\mathcal{D}|} \frac{1}{2}\left(EHH(x_{i-1}) + EHH(x_i)\right)g(x_{i-1},x_i) + \notag\\
&\negspace\sum_{i = 1}^{|\mathcal{U}|} \frac{1}{2}\left(EHH(x_{i-1}) + EHH(x_i)\right)g(x_{i-1},x_i)
\end{align}
If $iHH_A$ and $iHH_B$ are the iHHs for populations $A$ and $B$, then the 
(unstandardized) XPEHH is
\begin{equation}
\ln\left(\frac{iHH_A}{iHH_B}\right),
\end{equation}
and after genome-wide normalization we have
\begin{equation}
XPEHH = \frac{\ln\left(\frac{iHH_A}{iHH_B}\right) - E\Big[\ln\left(\frac{iHH_A}{iHH_B}\right)\Big]}{SD\Big[\ln\left(\frac{iHH_A}{iHH_B}\right)\Big]}.
\end{equation}

In practice, the sums in each of $iHH_A$ and $iHH_B$ (Equation \ref{eq:xpihh}) 
are truncated at $x_i$---the marker at which the EHH of the haplotypes {\it 
pooled across populations} is $EHH(x_i) < 0.05$.  Scaling of $g(x_{i-1},x_i)$ 
and handling of gaps is done as for iHS, and these parameters are definable 
on the {\tt selscan} command line.

\section{Performance}

Here we evaluate the performance of {\tt selscan} ({\tt https://github.com/szpiech/selscan}) for computing the iHS and XPEHH statistics. In addition, we compare performance on these statistics with the programs {\tt rehh} \cite[{\tt http://cran.r-project.org/package=rehh}]{GautierAndVitalis12}, {\tt ihs} \cite[]{VoightEtAl06} and {\tt xpehh} \cite[]{PickrellEtAl09}.  Both {\tt ihs} and {\tt xpehh} are available for download at {\tt http://hgdp.uchicago.edu/Software/}. All computations were run on a MacPro running OSX $10.8.5$ with two $2.4$ GHz $6-$core Intel Xeon processors with hyperthreading enabled.

\subsection{iHS}

For runtime evaluation of iHS calculations, we simulated a $4$ Mbp 
region of DNA with the program {\tt ms} \cite[]{Hudson02} and generated four 
independent data sets with varying numbers of sampled haplotypes ($\theta = 1600$ and $\rho = 1600$).
We sampled $250$ haplotypes ($9,625$ SNP loci), $500$ haplotypes  
($10,646$ SNP loci), $1,000$ haplotypes ($11,655$ SNP loci), and $2,000$ 
haplotypes ($12,724$ SNP loci).  We name these data sets IHS$250$, IHS$500$, 
IHS$1000$, IHS$2000$, respectively.  These data sets represent a densely typed 
region similar to next-generation sequencing data.  Although these
data sets are generated via strictly neutral processes, they serve the purpose of runtime
evaulation perfectly well.  We also use data from The $1000$ Genomes Project \cite[]{1000G12} Omni 
genotypes, calculating iHS scores at $22,147$ SNP loci on chromosome 22 across $102$ CEU 
individuals ($204$ haplotypes).  We name this data set CEU$22$.

Table \ref{tab:ihs} summarizes the runtimes of {\tt ihs}, {\tt rehh}, and {\tt selscan}.  We note that {\tt rehh} integrates
haplotype homozygosity over a physical map, whereas {\tt ihs} and {\tt selscan} integrate over a genetic map by default.  This does
not affect runtimes (data not shown), which are measured using genetic maps for {\tt ihs} and {\tt selscan}. Even operating on a single thread, {\tt selscan} calculates iHS scores at least an order of magnitude faster than {\tt ihs} and up to $1.8$x faster than {\tt rehh} for large data sets.

\begin{figure}[h] \begin{center}
\centerline{{\includegraphics[scale=0.65,angle=0]{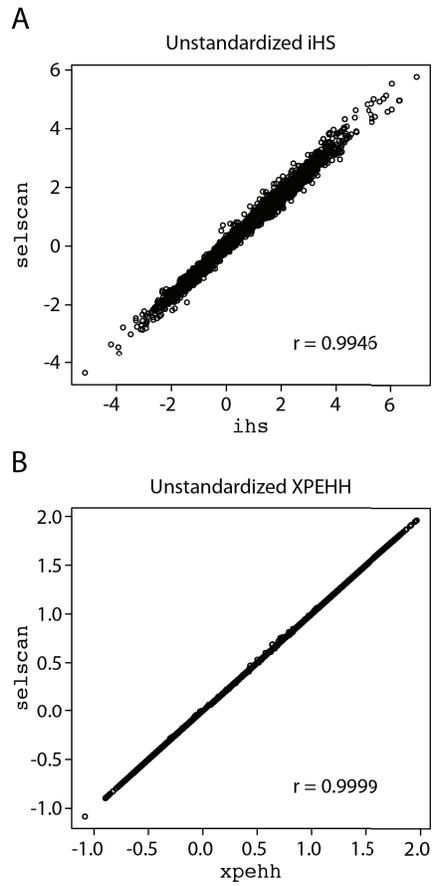}}}
\caption{(A) Unstandardized iHS scores calculated on the CEU$22$ data set for
  {\tt selscan} and {\tt ihs} (Pearson's $r=0.9946$) and (B)
  Unstandardized XPEHH scores calculated on the CEUYRI$22$ data set for
  {\tt selscan} and {\tt xpehh} (Pearson's $r=0.9999$)} \label{fig:cor} \end{center}
\end{figure}

We compare unstandardized iHS scores for the CEU$22$ data set using {\tt ihs} and {\tt selscan} and find excellent agreement (Figure \ref{fig:cor}A, Pearson's $r = 0.9946$).  The slight variance in scores between the two programs is likely due to an undocumented difference in the way {\tt ihs} calculates its scores (\cite{SabetiEtAl07} Supplemental Information), but the effect is negligible.  We also calculate unstandardized iHS scores for the CEU$22$ data set using {\tt rehh} and {\tt selscan} (using a physical map) and again find excellent agreement (Pearson's $r = 0.9953$).

\subsection{XPEHH}

For runtime evaluation of XPEHH calculations, we simulated a $4$ Mbp 
region of DNA with the program {\tt ms} \cite[]{Hudson02} with a
simple two population divergence model (time to divergence $t = 0.05$, $\theta = 1600$ and $\rho = 1600$)
and generated four independent data sets with varying numbers of sampled haplotypes.  We sampled
$250$ haplotypes ($125$ from each population, $12,920$ SNP loci), $500$ haplotypes ($250$ from each
population, $14,989$ SNP loci), $1,000$ haplotypes ($500$ from each population, $17,142$ SNP loci), and $2,000$ 
haplotypes ($1,000$ from each population, $19,567$ SNP loci).  We name these data sets XP$250$, XP$500$, 
XP$1000$, XP$2000$, respectively.  These data sets represent a densely typed 
region similar to next-generation sequencing data.  Although these
data sets are generated via strictly neutral processes, they serve the purpose of runtime
evaulation perfectly well.  We also use data from The $1000$ Genomes Project \cite[]{1000G12} Omni 
genotypes, calculating XPEHH scores at $22,147$ SNP loci on chromosome 22 across $102$ CEU 
individuals ($204$ haplotypes) and $105$ YRI individuals ($210$ haplotypes).  We name this data set 
CEUYRI$22$.

Table \ref{tab:xpehh} summarizes the runtimes of {\tt xpehh} and {\tt selscan}.  Even operating on a single thread,
{\tt selscan} tends to calculate XPEHH scores at least an order of magnitude faster than {\tt xpehh}. Figure \ref{fig:cor}B 
shows the correlation (Pearson's $r = 0.9999$) of CEUYRI$22$ unstandardized XPEHH scores between the two programs.

\section{Conclusions}

{\tt selscan} achieves a speed up of at least an order or magnitude over both {\tt ihs} and {\tt xpehh} and a speed up of nearly $2$x over {\tt rehh} for large data sets through
general optimizations of the calculations.  We also implement shared memory parallelism with multithreading to further speed up calculations on computers with multiple cores.  Since iHS and XPEHH attempt to calculate a score for each site in the data and each score can be calculated indpendently of the others, {\tt selscan} partitions the workload (sites at which to calculate a score) across threads, while maintaining each thread's access to the entire data set required to make the calculation.  

Additional empirical testing (data not shown) suggests that {\tt rehh},
{\tt ihs}, and {\tt selscan} (for both iHS and XPEHH calculations) are $O(ND^2)$, and {\tt xpehh} is $O(N^2D^2)$, where
$N$ is the number of haploid samples and $D$ is the SNP locus density.

Each of these statistics require phased haplotypes and a genetic or physical map as input data (TPED format) and missing genotypes must either be dropped or imputed.  Because of the speed improvements we have implented, we expect that {\tt selscan} will be a valuable tool for calculating EHH-based genome-wide scans for positive selection in very large genetic data sets, including whole genome sequencing and GWAS data, currently being generated for humans and other organisms.  {\tt selscan} will also allow for in-depth examination of the performance of these statistics under a wide range of parameters in large scale simulation studies.

\section{Acknowlegements}

The authors would like to thank Trevor Pemberton and Paul Verdu for assistance in testing the Windows binaries. This 
work was partially supported by the National Institutes of Health (grants P60MD006902, UL1RR024131, 1R21HG007233, 
1R21CA178706, 1R01HL117004-01, and 1R01HG007644) and a Sloan Foundation Research Fellowship (to R.D.H.).

\bibliographystyle{natbib}
\bibliography{Ref_ZAS}

\clearpage
\onecolumn

\begin{table}[h]
\begin{center}
\caption{Runtime performance (in seconds) of {\tt ihs}, {\tt rehh}, and {\tt selscan} for calculating unstandardized iHS for various data sets. Calculations running 
over 100,000 seconds were aborted. $^*${\tt rehh} integrates over a physical map instead of a genetic map.  Using a physical map does not affect {\tt selscan}'s runtime (data not shown). }
\label{tab:ihs} 
\begin{tabular}{|r|r|r|rrrrr|}
\hline
\multirow{2}{*}{Data Set} & \multirow{2}{*}{{\tt ihs}} & \multirow{2}{*}{{\tt rehh}$^*$} & \multicolumn{5}{|c|}{{\tt selscan}}\\ 
&&& threads $=1$ & $2$ & $4$ & $8$ & $16$\\
\hline
IHS$250$ & $19,275$ & $563$ & $618$ & $306$ & $162$ & $84$ & $58$\\
IHS$500$ & $45,547$ & $1,652$ & $1,554$ & $782$ & $399$ & $220$ & $150$\\
IHS$1000$ & $>100,000$ & $4,834$ & $4,018$ & $2,019$ & $1,040$ & $566$ & $380$\\
IHS$2000$ & $>100,000$ & $12,652$ & $7,054$ & $3,633$ & $1,869$ & $1,046$ & $752$\\
CEU$22$ & $19,434$ & $588$ & $353$ & $182$ & $93$ & $50$ & $33$\\
\hline
\end{tabular}
\end{center}
\end{table}


\begin{table}[h]
\begin{center}
\caption{Runtime performance (in seconds) of {\tt xpehh} and {\tt selscan} for calculating unstandardized XPEHH for various data sets. Calculations running 
over 100,000 seconds were aborted.} 
\label{tab:xpehh} 
\begin{tabular}{|r|r|rrrrr|}
\hline
\multirow{2}{*}{Data Set} & \multirow{2}{*}{{\tt xpehh}} & \multicolumn{5}{|c|}{{\tt selscan}}\\ 
&& threads $=1$ & $2$ & $4$ & $8$ & $16$\\
\hline
XP$250$ & $11,113$ & $287$ & $141$ & $71$ & $38$ & $25$\\
XP$500$ & $57,006$ & $766$ & $403$ & $194$ & $104$ & $67$\\
XP$1000$ & $>100,000$ & $2,037$ & $1,018$ & $515$ & $274$ & $180$\\
XP$2000$ & $>100,000$ & $5,683$ & $2,798$ & $1,471$ & $763$ & $493$\\
CEUYRI$22$ & $37,271$ & $578$ & $291$ & $150$ & $78$ & $52$\\
\hline
\end{tabular}
\end{center}
\end{table}

\end{document}